\documentclass[12pt]{article}

\usepackage{graphicx}
\usepackage{amsmath}
\usepackage{amssymb}

\allowdisplaybreaks

\begin{document}

\baselineskip=18.8pt plus 0.2pt minus 0.1pt

\makeatletter

\@addtoreset{equation}{section}
\renewcommand{\theequation}{\thesection.\arabic{equation}}
\renewcommand{\thefootnote}{\fnsymbol{footnote}}
\newcommand{\beq}{\begin{equation}}
\newcommand{\eeq}{\end{equation}}
\newcommand{\bea}{\begin{eqnarray}}
\newcommand{\eea}{\end{eqnarray}}
\newcommand{\nn}{\nonumber\\}
\newcommand{\hs}[1]{\hspace{#1}}
\newcommand{\vs}[1]{\vspace{#1}}
\newcommand{\Half}{\frac{1}{2}}
\newcommand{\p}{\partial}
\newcommand{\ol}{\overline}
\newcommand{\wt}[1]{\widetilde{#1}}
\newcommand{\ap}{\alpha'}
\newcommand{\bra}[1]{\left\langle  #1 \right\vert }
\newcommand{\ket}[1]{\left\vert #1 \right\rangle }
\newcommand{\vev}[1]{\left\langle  #1 \right\rangle }
\newcommand{\ul}[1]{\underline{#1}}

\newcommand{\hook}{\raisebox{-0.35ex}{\makebox[0.6em][r]
{\scriptsize $-$}}\hspace{-0.15em}\raisebox{0.25ex}{\makebox[0.4em][l]{\tiny
 $|$}}}

\makeatother

\begin{titlepage}
\title{
\vspace{1cm}
On quantum numbers for Rarita-Schwinger fields
}
\author{Yoji Michishita
\thanks{
{\tt michishita@edu.kagoshima-u.ac.jp}
}
\\[7pt]
{\it Department of Physics, Faculty of Education, Kagoshima University}\\
{\it Kagoshima, 890-0065, Japan}
}

\date{\normalsize October, 2018}
\maketitle
\thispagestyle{empty}

\begin{abstract}
\normalsize
We consider first order linear operators commuting with the operator 
appearing in the linearized equation of motion of Rarita-Schwinger fields
which comes directly from the action.
First we consider a simplified operator giving an equation equivalent to the original equation,
and classify first order operators commuting with it in four dimensions. 
In general such operators are symmetry operators of the original operator,
but we find that some of them commute with it.
We extend this result in four dimensions to arbitrary dimensions
and give first order commuting operators constructed of odd rank Killing-Yano and 
even rank closed conformal Killing-Yano tensors with additional conditions.
\end{abstract}
\end{titlepage}

\clearpage
\section{Introduction}

It is important to obtain solutions to the linearized equations of motion of various fields
in given background geometries. Schematically the equations are in the form of $M\phi=0$,
where $\phi$ is the field and $M$ is some operator which is second order in derivative operator for boson fields,
and first order for fermion fields. It is useful to know if there exists a linear operator commuting with $M$
because it gives a quantum number for classifying solutions to $M\phi=0$.
If $M$ is the one coming from the action of $\phi$ directly i.e. if the Lagrangian is given by
$\mathcal{L}\sim \phi M\phi$, such operators are also useful to compute Euclidean path integrals.

For spin $1/2$ fields $M$ is the Dirac operator, and first order linear operators commuting with it
(or more generally symmetry operators) are
completely classified\cite{ms79,bc96,bk04,ckk11}: They are constructed of
Killing-Yano forms of odd rank and closed conformal Killing-Yano (CCKY) forms of even rank.
For some extensions of this result see \cite{aeov08, hkwy10}.

In this paper we consider this issue for Rarita-Schwinger field $\psi_\mu$.
(For some related discussions see \cite{so95,so97,ae17}.)
We only consider the case where $\psi_\mu$ is a Dirac vector-spinor, and
in section 2 and 3 we only consider the case where the spacetime dimension $D$ is even.
In section 4 $D$ is set 4,  and in section 5 we do not impose any restriction on $D$. 
We put no assumption on the signature of the background metric.
The linearized equation of motion derived directly from the action is $M_\mu{}^\nu\psi_\nu=0$, and 
$M_\mu{}^\nu$ is given by
\beq
M_\mu{}^\nu=\Gamma_\mu{}^{\lambda\nu}\mathcal{D}_\lambda-m\Gamma_\mu{}^\nu,
\eeq
where $m$ is the mass parameter, and $\mathcal{D}_\mu=\mathcal{D}_\mu(\Gamma,\omega)$
is the covariant derivative with respect to the spin connection and the affine connection.
However we found that the calculation is quite complicated if we use this operator directly, and 
to simplify the situation we consider the following $\hat{M}_\mu{}^\nu$ instead:
\beq
\hat{M}_\mu{}^\nu =
\delta_\mu{}^\nu\Gamma^\lambda\mathcal{D}_\lambda - \Gamma^\nu\mathcal{D}_\mu
 +\frac{m}{D-2}\Gamma_\mu{}^\nu+\frac{D-1}{D-2}m\delta_\mu{}^\nu.
\eeq
It is easy to show that
$\hat{M}_\mu{}^\nu\psi_\nu=0$ is equivalent to $M_\mu{}^\nu\psi_\nu=0$.
Some operators commuting with $\hat{M}_\mu{}^\nu$ also commute with $M_\mu{}^\nu$,
and we give such examples constructed of Killing-Yano forms of odd rank and CCKY forms of even rank,
as in the case of the Dirac operator.

In section 2 
we give conditions for first order operators commuting with $\hat{M}_\mu{}^\nu$ for even $D$, and
in section 3 we consider how to solve the conditions.
Since it is not easy to obtain full solution to the conditions for general $D$, in section 4 we set $D=4$ and
solve the conditions in a brute-force manner by using symbolic manipulation program Mathematica
and the package GAMMA\cite{gran01}.
It turns out that the solution for massive case also commutes with $M_\mu{}^\nu$.
In section 5, we try to extend this result for $D=4$ to arbitrary $D$.
We find operators commuting with $M_\mu{}^\nu$ constructed of 
odd rank Killing-Yano tensors and even rank CCKY tensors.
This situation is similar to the case of the Dirac operator, but
unlike that case we have to impose additional conditions for those tensors.
Section 6 contains some discussions, and in Appendix A we briefly summarize 
properties of Killing-Yano tensors and CCKY tensors used in the calculation.
We mainly use gamma matrix notation which is more familiar to physicists than that used in 
the literature as \cite{bc96,bk04,ckk11} and suitable for calculations using Mathematica.
In Appendix B we give some of our results in the notation of \cite{bc96,bk04,ckk11}
for readers' convenience.

\section{Preliminaries}

We consider Rarita-Schwinger field $\psi_\mu$ in $D$-dimensional space.
$\psi_\mu$ is a Dirac vector-spinor, and we do not consider Majorana or Weyl case.
Since $\psi_\mu$ is not dynamical for $D\le 2$, we consider cases of $D\ge 3$.
No assumption on the signature of the background metric $g_{\mu\nu}$ is made,
and we do not impose background equation of motion on $g_{\mu\nu}$ .

The linearized equation of motion for $\psi_\mu$ is given by the following form,
which comes directly from the action principle:
\beq
(\Gamma^{\mu\lambda\nu}\mathcal{D}_\lambda-m\Gamma^{\mu\nu})\psi_\nu=0.
\label{rseq}
\eeq
This equation is expressed as $M_\mu{}^\nu\psi_\nu=0$, where 
$M_\mu{}^\nu$ is 
\beq
M_\mu{}^\nu=\Gamma_\mu{}^{\lambda\nu}\mathcal{D}_\lambda-m\Gamma_\mu{}^\nu.
\eeq
The equation is also expressed as $\hat{M}_\mu{}^\nu\psi_\nu=0$, where
\bea
\hat{M}_\mu{}^\nu & = &
 \Big(\delta_\mu{}^\rho-\frac{1}{D-2}\Gamma_\mu\Gamma^\rho\Big)M_\rho{}^\nu
\nn & = & \delta_\mu{}^\nu\Gamma^\lambda\mathcal{D}_\lambda
 -\Gamma^\nu\mathcal{D}_\mu
 +\frac{m}{D-2}\Gamma_\mu{}^\nu+\frac{D-1}{D-2}m\delta_\mu{}^\nu.
\eea
Conversely $M_\mu{}^\nu$ is given by $\hat{M}_\mu{}^\nu$ as follows:
\beq
M_\mu{}^\nu =
 \Big(\delta_\mu{}^\rho-\frac{1}{2}\Gamma_\mu\Gamma^\rho\Big)\hat{M}_\rho{}^\nu.
\eeq
It is well-known that two mutually commuting diagonalizable matrices are simultaneously diagonalizable,
and it is important to give operators commuting with $M_\mu{}^\nu$ or $\hat{M}_\mu{}^\nu$.
If the background is Euclidean, $iM_\mu{}^\nu$ is a hermitian operator if $m$ is pure imaginary, and 
its diagonalizability is assured. (We do not take complex conjugation in the following calculation, and it is possible
to regard $m$ as complex formally.)
However $i\hat{M}_\mu{}^\nu$ is not hermitian, and
it is not immediately clear if $\hat{M}_\mu{}^\nu$ is diagonalizable or not. 
Although $\hat{M}_\mu{}^\nu$ has such issues, ranks of gamma matrices in $\hat{M}_\mu{}^\nu$
is lower than those of $M_\mu{}^\nu$ and it makes calculations easier. 
Furthermore solutions to
the equation of motion belong to the eigenspace of eigenvalue 0, which is
common to $M_\mu{}^\nu$ and $\hat{M}_\mu{}^\nu$.
Therefore in this section and the next two sections we consider first order operator 
$O_\mu{}^\nu$ commuting with $\hat{M}_\mu{}^\nu$:
\beq
[O,\hat{M}]_\mu{}^\nu\psi_\nu
=(O_\mu{}^\rho \hat{M}_\rho{}^\nu-\hat{M}_\mu{}^\rho O_\rho{}^\nu)\psi_\nu=0.
\eeq
This means that
\beq
\Big(\delta_\mu{}^\lambda-\frac{1}{2}\Gamma_\mu\Gamma^\lambda\Big)
O_\lambda{}^\sigma\Big(\delta_\sigma{}^\rho-\frac{1}{D-2}\Gamma_\sigma\Gamma^\rho\Big)
M_\rho{}^\nu\psi_\nu - M_\mu{}^\rho O_\rho{}^\nu\psi_\nu
=0.
\eeq
Therefore $O_\mu{}^\nu$ is a symmetry operator for $M_\mu{}^\nu$.
If $O_\mu{}^\nu$ is diagonalizable in the eigenspace of eigenvalue 0 then we can use it to classify
the solutions. Even if it is not diagonalizable, there exists an eigenvector common to $\hat{M}_\mu{}^\nu$ and $O_\mu{}^\nu$.

If $O_\mu{}^\nu$ satisfies
\beq
\Big(\delta_\mu{}^\lambda-\frac{1}{2}\Gamma_\mu\Gamma^\lambda\Big)
O_\lambda{}^\sigma\Big(\delta_\sigma{}^\nu-\frac{1}{D-2}\Gamma_\sigma\Gamma^\nu\Big)
= O_\mu{}^\nu,
\label{mhattom}
\eeq
then $O_\mu{}^\nu$ commutes also with $M_\mu{}^\nu$. Then simultaneous diagonalizability of the hermitian part and
the antihermitian part of $O_\mu{}^\nu$ with $M_\mu{}^\nu$ is assured for Euclidean backgrounds.
We will see examples of such operators later.

In the case of spin 1/2 fields, the mass term is proportional to the unit matrix. 
Therefore it makes no difference to commuting operators
(except that Weyl condition can be imposed further in the massless case.)
However for spin 3/2 fields the mass term is not trivial, and the massive case may be more restrictive
than the massless case.

$O_\mu{}^\nu$ is decomposed into derivative part $P_\mu{}^\nu$ and 
nonderivative part $Q_\mu{}^\nu$:
\beq
O_\mu{}^\nu = P_\mu{}^\nu + Q_\mu{}^\nu,
\eeq
\beq
P_\mu{}^\nu = \sum_{n=0}^D\Gamma^{\lambda_1\dots\lambda_n}
 F_\mu{}^\nu{}_{\lambda_1\dots\lambda_n}{}^\sigma\mathcal{D}_\sigma,
\eeq
\beq
Q_\mu{}^\nu = \sum_{n=0}^D\Gamma^{\lambda_1\dots\lambda_n}
 f_\mu{}^\nu{}_{\lambda_1\dots\lambda_n}.
\eeq
If $D$ is odd, then we can restrict the range of the dummy index $n$ in the above expression 
of $P_\mu{}^\nu$ and $Q_\mu{}^\nu$ to $0\le n\le (D-1)/2$, or to even integers. 
In this section and the next section we only consider even $D$ case.

The second order part in $[O,\hat{M}]_\mu{}^\nu$ comes from the commutator between $P_\mu{}^\nu$ and
$\delta_\mu{}^\nu\Gamma^\lambda\mathcal{D}_\lambda-\Gamma^\nu\mathcal{D}_\mu$:
\bea
\lefteqn{\Big\{P_\mu{}^\rho(\delta_\rho{}^\nu\Gamma^\lambda\mathcal{D}_\lambda
 -\Gamma^\nu\mathcal{D}_\rho)
-(\delta_\mu{}^\rho\Gamma^\lambda\mathcal{D}_\lambda
 -\Gamma^\rho\mathcal{D}_\mu)P_\rho{}^\nu
\Big\}\psi_\nu}
\nn & = &
 \sum_{n=0}^{D-1}(n+1)\Big\{F_\mu{}^\nu{}_{\lambda_1\dots\lambda_n}{}^{\sigma\rho}
 -(-1)^nF_\mu{}^\nu{}_{\lambda_1\dots\lambda_n}{}^{\rho\sigma}
\nn & &
 -F_\mu{}^\sigma{}_{\lambda_1\dots\lambda_n}{}^{\nu\rho}
 +(-1)^n\delta_\mu{}^\rho F_\tau{}^\nu{}_{\lambda_1\dots\lambda_n}{}^{\tau\sigma}\Big\}
 \Gamma^{\lambda_1\dots\lambda_n}\mathcal{D}_\rho\mathcal{D}_\sigma\psi_\nu
\nn & &
+\sum_{n=1}^D \Big\{F_\mu{}^\nu{}_{\lambda_1\dots\lambda_{n-1}}{}^\rho\delta_{\lambda_n}{}^\sigma
 +(-1)^nF_\mu{}^\nu{}_{\lambda_1\dots\lambda_{n-1}}{}^\sigma\delta_{\lambda_n}{}^\rho
\nn & &
 -F_\mu{}^\sigma{}_{\lambda_1\dots\lambda_{n-1}}{}^\rho\delta_{\lambda_n}{}^\nu
 -(-1)^nF_{\lambda_n}{}^\nu{}_{\lambda_1\dots\lambda_{n-1}}{}^\sigma\delta_\mu{}^\rho
\Big\}
\Gamma^{\lambda_1\dots\lambda_n}\mathcal{D}_\rho\mathcal{D}_\sigma\psi_\nu
\nn &  &
+\sum_{n=0}^{D-1}(n+1)\Big\{
 -\nabla_\mu F_\sigma{}^{\nu\sigma}{}_{\lambda_1\dots\lambda_n}{}^\rho
 +(-1)^n \nabla_\sigma F_\mu{}^\nu{}_{\lambda_1\dots\lambda_n}{}^{\sigma\rho}
\Big\}\mathcal{D}_\rho\psi_\nu
\nn & &
+\sum_{n=1}^D \Big\{
 -\nabla_\mu F_{\lambda_1}{}^\nu{}_{\lambda_2\dots\lambda_n}{}^\rho
 -(-1)^n \nabla_{\lambda_n} F_\mu{}^\nu{}_{\lambda_1\dots\lambda_{n-1}}{}^\rho
\Big\}\mathcal{D}_\rho\psi_\nu,
\label{secondder}
\eea
where $\nabla_\mu$ is the covariant derivative with respect to the affine connection.
Since $\mathcal{D}_{\rho}\mathcal{D}_{\sigma}\psi_\nu$ is decomposed into 
the antisymmetric part and the symmetric part:
\beq
\mathcal{D}_{\rho}\mathcal{D}_{\sigma}\psi_\nu
=
\mathcal{D}_{[\rho}\mathcal{D}_{\sigma]}\psi_\nu
+\mathcal{D}_{(\rho}\mathcal{D}_{\sigma)}\psi_\nu,
\eeq
and the antisymmetric part can be rewritten as the form without derivative operator:
\bea
\Gamma^{\lambda_1\dots\lambda_n}
\mathcal{D}_{[\rho}\mathcal{D}_{\sigma]}\psi_\nu
 & =  &
\Gamma^{\lambda_1\dots\lambda_n}\left(\frac{1}{2}R_{\rho\sigma\nu}{}^\tau\psi_\tau
 +\frac{1}{8}R_{\rho\sigma}{}^{ab}\Gamma_{ab}\psi_\nu\right)
\nn & = & 
 \frac{1}{2}R_{\rho\sigma\nu}{}^\tau\Gamma^{\lambda_1\dots\lambda_n}\psi_\tau
 +\frac{1}{8}R_{\rho\sigma\lambda_{n+1}\lambda_{n+2}}\Gamma^{\lambda_1\dots\lambda_{n+2}}\psi_\nu
\nn & &
 +\frac{n}{4}R_{\rho\sigma}{}^{[\lambda_n}{}_\tau\Gamma^{\lambda_1\dots\lambda_{n-1}]\tau}\psi_\nu
 -\frac{1}{8}n(n-1)R_{\rho\sigma}{}^{[\lambda_{n-1}\lambda_n}\Gamma^{\lambda_1\dots\lambda_{n-2}]}\psi_\nu,
\label{comcd}
\eea
then the coefficients of the symmetric part in \eqref{secondder} must vanish.
From this condition, for odd $n$ we obtain
\bea
0 & = & (n+1)\Big(2F_\mu{}^\nu{}_{\lambda_1\dots\lambda_n}{}^{\rho\sigma}
 -F_\mu{}^\rho{}_{\lambda_1\dots\lambda_n}{}^{\nu\sigma}
 -\delta_\mu{}^\rho F_\tau{}^\nu{}_{\lambda_1\dots\lambda_n}{}^{\tau\sigma}\Big)
\nn & &
 -F_\mu{}^\rho{}_{[\lambda_1\dots\lambda_{n-1}}{}^\sigma\delta_{\lambda_n]}{}^\nu
 +F_{[\lambda_n}{}^\nu{}_{\lambda_1\dots\lambda_{n-1}]}{}^\sigma\delta_\mu{}^\rho
\nn & &
 +(\rho\leftrightarrow\sigma),
\label{oddF0}
\eea
and for even $n$,
\bea
0 & = & (n+1)\Big(
 -F_\mu{}^\rho{}_{\lambda_1\dots\lambda_n}{}^{\nu\sigma}
 +\delta_\mu{}^\rho F_\tau{}^\nu{}_{\lambda_1\dots\lambda_n}{}^{\tau\sigma}\Big)
 +2F_\mu{}^\nu{}_{[\lambda_1\dots\lambda_{n-1}}{}^\sigma\delta_{\lambda_n]}{}^\rho
\nn & &
 -F_\mu{}^\rho{}_{[\lambda_1\dots\lambda_{n-1}}{}^\sigma\delta_{\lambda_n]}{}^\nu
 -F_{[\lambda_n}{}^\nu{}_{\lambda_1\dots\lambda_{n-1}]}{}^\sigma\delta_\mu{}^\rho
\nn & &
 +(\rho\leftrightarrow\sigma).
\label{evenF0}
\eea
For $n=0$, expressions as $F_\mu{}^\rho{}_{[\lambda_1\dots\lambda_{n-1}}{}^\sigma\delta_{\lambda_n]}{}^\nu$
in the above do not make sense, and here and in the following it is understood that such terms are just omitted. 

Next, we require that the first order part of $[O,\hat{M}]_\mu{}^\nu\psi_\nu$ vanishes.
Then we obtain the following relations:
For odd $n$,
\bea
0 & = & 
(n+1)\Big(
 2f_\mu{}^\nu{}_{\lambda_1\dots\lambda_n}{}^\sigma
 -f_\mu{}^\sigma{}_{\lambda_1\dots\lambda_n}{}^\nu
 -\delta_\mu{}^\sigma f_\tau{}^\nu{}_{\lambda_1\dots\lambda_n}{}^\tau\Big)
\nn & &
 -\delta_{[\lambda_n}{}^\nu f_{|\mu|}{}^\sigma{}_{\lambda_1\dots\lambda_{n-1}]}
 +\delta_\mu{}^\sigma f_{[\lambda_n}{}^\nu{}_{\lambda_1\dots\lambda_{n-1}]}
\nn & &
+(n+1)\Big(
 \nabla_\mu F_\rho{}^{\nu\rho}{}_{\lambda_1\dots\lambda_n}{}^\sigma
 +\nabla_\rho F_\mu{}^\nu{}_{\lambda_1\dots\lambda_n}{}^{\rho\sigma}
\Big)
\nn & &
 -\nabla_{[\lambda_n} F_{|\mu|}{}^\nu{}_{\lambda_1\dots\lambda_{n-1}]}{}^\sigma
 +\nabla_\mu F_{[\lambda_1}{}^\nu{}_{\lambda_2\dots\lambda_n]}{}^\sigma
\nn & &
+\frac{m}{D-2}\Big\{
 (n+1)(n+2)\Big(
 F_\rho{}^\nu{}_\mu{}^\rho{}_{\lambda_1\dots\lambda_n}{}^\sigma
 -F_{\mu\rho}{}^{\rho\nu}{}_{\lambda_1\dots\lambda_n}{}^\sigma
\Big)
\nn & &
+n\Big(
 F_{[\lambda_1}{}^\nu{}_{|\mu|\lambda_2\dots\lambda_n]}{}^\sigma
 -F_{\mu[\lambda_1}{}^\nu{}_{\lambda_2\dots\lambda_n]}{}^\sigma
 -g_{\mu[\lambda_1}F_{|\rho|}{}^{\nu\rho}{}_{\lambda_2\dots\lambda_n]}{}^\sigma
 +\delta^\nu{}_{[\lambda_1}F_{|\mu|}{}^\rho{}_{\lambda_2\dots\lambda_n]\rho}{}^\sigma
\Big)
\nn & &
-g_{\mu[\lambda_1}F_{\lambda_2}{}^\nu{}_{\lambda_3\dots\lambda_n]}{}^\sigma
 -\delta^\nu{}_{[\lambda_1}F_{|\mu|\lambda_2\dots\lambda_n]}{}^\sigma
\Big\}.
\label{oddf0}
\eea
For even $n$,
\bea
0 & = & 
-(n+1)\Big(f_\mu{}^\sigma{}_{\lambda_1\dots\lambda_n}{}^\nu
 -\delta_\mu{}^\sigma f_\tau{}^\nu{}_{\lambda_1\dots\lambda_n}{}^\tau\Big)
\nn & &
+2\delta_{[\lambda_n}{}^\sigma f_{|\mu|}{}^\nu{}_{\lambda_1\dots\lambda_{n-1}]}
 -\delta_{[\lambda_n}{}^\nu f_{|\mu|}{}^\sigma{}_{\lambda_1\dots\lambda_{n-1}]}
 -\delta_\mu{}^\sigma f_{[\lambda_n}{}^\nu{}_{\lambda_1\dots\lambda_{n-1}]}
\nn & &
+(n+1)\Big(
 \nabla_\mu F_\rho{}^{\nu\rho}{}_{\lambda_1\dots\lambda_n}{}^\sigma
 -\nabla_\rho F_\mu{}^\nu{}_{\lambda_1\dots\lambda_n}{}^{\rho\sigma}
\Big)
\nn & &
 +\nabla_{[\lambda_n} F_{|\mu|}{}^\nu{}_{\lambda_1\dots\lambda_{n-1}]}{}^\sigma
 +\nabla_\mu F_{[\lambda_1}{}^\nu{}_{\lambda_2\dots\lambda_n]}{}^\sigma
\nn & &
+\frac{m}{D-2}\Big\{
 (n+1)(n+2)\Big(
 F_\rho{}^\nu{}_\mu{}^\rho{}_{\lambda_1\dots\lambda_n}{}^\sigma
 -F_{\mu\rho}{}^{\rho\nu}{}_{\lambda_1\dots\lambda_n}{}^\sigma
\Big)
\nn & &
+n\Big(
 F_{[\lambda_1}{}^\nu{}_{|\mu|\lambda_2\dots\lambda_n]}{}^\sigma
 -F_{\mu[\lambda_1}{}^\nu{}_{\lambda_2\dots\lambda_n]}{}^\sigma
 -g_{\mu[\lambda_1}F_{|\rho|}{}^{\nu\rho}{}_{\lambda_2\dots\lambda_n]}{}^\sigma
 -\delta^\nu{}_{[\lambda_1}F_{|\mu|}{}^\rho{}_{\lambda_2\dots\lambda_n]\rho}{}^\sigma
\Big)
\nn & &
-g_{\mu[\lambda_1}F_{\lambda_2}{}^\nu{}_{\lambda_3\dots\lambda_n]}{}^\sigma
 -\delta^\nu{}_{[\lambda_1}F_{|\mu|\lambda_2\dots\lambda_n]}{}^\sigma
\Big\}.
\label{evenf0}
\eea

Finally we require that the zeroth order part of $[O,\hat{M}]_\mu{}^\nu\psi_\nu$ vanishes.
For odd $n$,
\bea
0 & = & (n+1)\Big(
 \nabla_\mu f_\rho{}^{\nu\rho}{}_{\lambda_1\dots\lambda_n}
 +\nabla_\rho f_\mu{}^\nu{}_{\lambda_1\dots\lambda_n}{}^\rho
\Big)
\nn & &
 +\nabla_\mu f_{[\lambda_1}{}^\nu{}_{\lambda_2\dots\lambda_n]}
 -\nabla_{[\lambda_n}f_{|\mu|}{}^\nu{}_{\lambda_1\dots\lambda_{n-1}]}
\nn & &
+F_\mu{}^\tau{}_{[\lambda_1\dots\lambda_{n-1}}{}^\rho\delta_{\lambda_n]}{}^\sigma
 R_{\rho\sigma\tau}{}^\nu
\nn & &
+\frac{1}{2}nF_\mu{}^\nu{}_{[\tau_1\dots\tau_{n-1}}{}^\rho\delta_{\tau_n]}{}^\sigma
 R_{\rho\sigma}{}^{\tau_n}{}_{[\lambda_n}\delta_{\lambda_1}{}^{\tau_1}\dots\delta_{\lambda_{n-1}]}{}^{\tau_{n-1}}
\nn & &
 -\frac{1}{4}(n+1)(n+2)F_\mu{}^\nu{}_{[\lambda_1\dots\lambda_n\tau}{}^\rho\delta_{\phi]}{}^\sigma
 R_{\rho\sigma}{}^{\tau\phi}
\nn & &
 -\frac{1}{2}(n+1)\Big(F_\mu{}^\sigma{}_{\lambda_1\dots\lambda_n}{}^{\tau\rho}
 +\delta_\mu{}^\rho F_\phi{}^\tau{}_{\lambda_1\dots\lambda_n}{}^{\phi\sigma}
\Big)R_{\rho\sigma\tau}{}^\nu
\nn & &
 -\frac{1}{4}n(n+1)\Big(F_\mu{}^\sigma{}_{[\lambda_1\dots\lambda_{n-1}|\tau}{}^{\nu\rho}
 +\delta_\mu{}^\rho F_\phi{}^\nu{}_{[\lambda_1\dots\lambda_{n-1}|\tau}{}^{\phi\sigma}
\Big)R_{\rho\sigma|}{}^\tau{}_{\lambda_n]}
\nn & &
 +\frac{1}{8}(n+1)(n+2)(n+3)\Big(F_\mu{}^\sigma{}_{\lambda_1\dots\lambda_n\tau\phi}{}^{\nu\rho}
 +\delta_\mu{}^\rho F_\kappa{}^\nu{}_{\lambda_1\dots\lambda_n\tau\phi}{}^{\kappa\sigma}
\Big)R_{\rho\sigma}{}^{\tau\phi}
\nn & &
 -\frac{1}{8}(n-1)\Big(F_\mu{}^\sigma{}_{[\lambda_1\dots\lambda_{n-2}}{}^{\nu\rho}
 +\delta_\mu{}^\rho F_\phi{}^\nu{}_{[\lambda_1\dots\lambda_{n-2}}{}^{\phi\sigma}
\Big)R_{|\rho\sigma|\lambda_{n-1}\lambda_n]}
\nn & &
-\frac{1}{2}\Big(\delta_{[\lambda_n}{}^\tau F_{|\mu|}{}^\sigma{}_{\lambda_1\dots\lambda_{n-1}]}{}^\rho
 -\delta_\mu{}^\rho F_{[\lambda_n}{}^\tau{}_{\lambda_1\dots\lambda_{n-1}]}{}^\sigma
\Big)R_{\rho\sigma\tau}{}^\nu
\nn & &
-\frac{1}{4}n\Big(\delta_{[\tau_n}{}^\nu F_{|\mu|}{}^\sigma{}_{\tau_1\dots\tau_{n-1}]}{}^\rho
 -\delta_\mu{}^\rho F_{[\tau_n}{}^\nu{}_{\tau_1\dots\tau_{n-1}]}{}^\sigma
\Big)R_{\rho\sigma}{}^{\tau_n}{}_{[\lambda_n}\delta_{\lambda_1}{}^{\tau_1}\dots\delta_{\lambda_{n-1}]}{}^{\tau_{n-1}}
\nn & &
+\frac{1}{8}(n+1)(n+2)\Big(\delta_{[\phi}{}^\nu F_{|\mu|}{}^\sigma{}_{\lambda_1\dots\lambda_n\tau]}{}^\rho
 -\delta_\mu{}^\rho F_{[\phi}{}^\nu{}_{\lambda_1\dots\lambda_n\tau]}{}^\sigma
\Big)R_{\rho\sigma}{}^{\tau\phi}
\nn & &
-\frac{1}{8}\Big(\delta_{[\lambda_{n-2}}{}^\nu F_{|\mu|}{}^\sigma{}_{\lambda_1\dots\lambda_{n-3}}{}^\rho
 -\delta_\mu{}^\rho F_{[\lambda_{n-2}}{}^\nu{}_{\lambda_1\dots\lambda_{n-3}}{}^\sigma
\Big)R_{|\rho\sigma|\lambda_{n-1}\lambda_n]}
\nn & &
+\frac{m}{D-2}\Big\{
 (n+1)(n+2)\Big(
 f_\rho{}^\nu{}_\mu{}^\rho{}_{\lambda_1\dots\lambda_n}
 -f_{\mu\rho}{}^{\rho\nu}{}_{\lambda_1\dots\lambda_n}
\Big)
\nn & &
+n\Big(
 f_{[\lambda_1}{}^\nu{}_{|\mu|\lambda_2\dots\lambda_n]}
 -f_{\mu[\lambda_1}{}^\nu{}_{\lambda_2\dots\lambda_n]}
 -g_{\mu[\lambda_1}f_{|\rho|}{}^{\nu\rho}{}_{\lambda_2\dots\lambda_n]}
 +\delta^\nu{}_{[\lambda_1}f_{|\mu|}{}^\rho{}_{\lambda_2\dots\lambda_n]\rho}
\Big)
\nn & &
-g_{\mu[\lambda_1}f_{\lambda_2}{}^\nu{}_{\lambda_3\dots\lambda_n]}
 -\delta^\nu{}_{[\lambda_1}f_{|\mu|\lambda_2\dots\lambda_n]}
\Big\}.
\label{oddD0}
\eea

For even $n$,
\bea
0 & = & (n+1)\Big(
 \nabla_\mu f_\rho{}^{\nu\rho}{}_{\lambda_1\dots\lambda_n}
 -\nabla_\rho f_\mu{}^\nu{}_{\lambda_1\dots\lambda_n}{}^\rho
\Big)
\nn & &
 +\nabla_\mu f_{[\lambda_1}{}^\nu{}_{\lambda_2\dots\lambda_n]}
 +\nabla_{[\lambda_n}f_{|\mu|}{}^\nu{}_{\lambda_1\dots\lambda_{n-1}]}
\nn & &
 +(n+1)F_\mu{}^\tau{}_{\lambda_1\dots\lambda_n}{}^{\sigma\rho}R_{\rho\sigma\tau}{}^\nu
 +\frac{1}{2}n(n+1)F_\mu{}^\nu{}_{[\lambda_1\dots\lambda_{n-1}|\tau}{}^{\sigma\rho}
 R_{\rho\sigma|}{}^\tau{}_{\lambda_n]}
\nn & &
 +\frac{1}{4}(n-1)F_\mu{}^\nu{}_{[\lambda_1\dots\lambda_{n-2}}{}^{\sigma\rho}
 R_{|\rho\sigma|\lambda_{n-1}\lambda_n]}
\nn & &
 -\frac{1}{2}(n+1)\Big(F_\mu{}^\sigma{}_{\lambda_1\dots\lambda_n}{}^{\tau\rho}
 -\delta_\mu{}^\rho F_\phi{}^\tau{}_{\lambda_1\dots\lambda_n}{}^{\phi\sigma}
\Big)R_{\rho\sigma\tau}{}^\nu
\nn & &
 -\frac{1}{4}n(n+1)\Big(F_\mu{}^\sigma{}_{[\lambda_1\dots\lambda_{n-1}|\tau}{}^{\nu\rho}
 -\delta_\mu{}^\rho F_\phi{}^\nu{}_{[\lambda_1\dots\lambda_{n-1}|\tau}{}^{\phi\sigma}
\Big)R_{\rho\sigma|}{}^\tau{}_{\lambda_n]}
\nn & &
 +\frac{1}{8}(n+1)(n+2)(n+3)\Big(F_\mu{}^\sigma{}_{\lambda_1\dots\lambda_n\tau\phi}{}^{\nu\rho}
 -\delta_\mu{}^\rho F_\kappa{}^\nu{}_{\lambda_1\dots\lambda_n\tau\phi}{}^{\kappa\sigma}
\Big)R_{\rho\sigma}{}^{\tau\phi}
\nn & &
 -\frac{1}{8}(n-1)\Big(F_\mu{}^\sigma{}_{[\lambda_1\dots\lambda_{n-2}}{}^{\nu\rho}
 -\delta_\mu{}^\rho F_\phi{}^\nu{}_{[\lambda_1\dots\lambda_{n-2}}{}^{\phi\sigma}
\Big)R_{|\rho\sigma|\lambda_{n-1}\lambda_n]}
\nn & &
-\frac{1}{2}\Big(\delta_{[\lambda_n}{}^\tau F_{|\mu|}{}^\sigma{}_{\lambda_1\dots\lambda_{n-1}]}{}^\rho
 +\delta_\mu{}^\rho F_{[\lambda_n}{}^\tau{}_{\lambda_1\dots\lambda_{n-1}]}{}^\sigma
\Big)R_{\rho\sigma\tau}{}^\nu
\nn & &
-\frac{1}{4}n\Big(\delta_{[\tau_n}{}^\nu F_{|\mu|}{}^\sigma{}_{\tau_1\dots\tau_{n-1}]}{}^\rho
 +\delta_\mu{}^\rho F_{[\tau_n}{}^\nu{}_{\tau_1\dots\tau_{n-1}]}{}^\sigma
\Big)R_{\rho\sigma}{}^{\tau_n}{}_{[\lambda_n}\delta_{\lambda_1}{}^{\tau_1}\dots\delta_{\lambda_{n-1}]}{}^{\tau_{n-1}}
\nn & &
+\frac{1}{8}(n+1)(n+2)\Big(\delta_{[\phi}{}^\nu F_{|\mu|}{}^\sigma{}_{\lambda_1\dots\lambda_n\tau]}{}^\rho
 +\delta_\mu{}^\rho F_{[\phi}{}^\nu{}_{\lambda_1\dots\lambda_n\tau]}{}^\sigma
\Big)R_{\rho\sigma}{}^{\tau\phi}
\nn & &
-\frac{1}{8}\Big(\delta_{[\lambda_{n-2}}{}^\nu F_{|\mu|}{}^\sigma{}_{\lambda_1\dots\lambda_{n-3}}{}^\rho
 +\delta_\mu{}^\rho F_{[\lambda_{n-2}}{}^\nu{}_{\lambda_1\dots\lambda_{n-3}}{}^\sigma
\Big)R_{|\rho\sigma|\lambda_{n-1}\lambda_n]}
\nn & &
+\frac{m}{D-2}\Big\{
 (n+1)(n+2)\Big(
 f_\rho{}^\nu{}_\mu{}^\rho{}_{\lambda_1\dots\lambda_n}
 -f_{\mu\rho}{}^{\rho\nu}{}_{\lambda_1\dots\lambda_n}
\Big)
\nn & &
+n\Big(
 f_{[\lambda_1}{}^\nu{}_{|\mu|\lambda_2\dots\lambda_n]}
 -f_{\mu[\lambda_1}{}^\nu{}_{\lambda_2\dots\lambda_n]}
 -g_{\mu[\lambda_1}f_{|\rho|}{}^{\nu\rho}{}_{\lambda_2\dots\lambda_n]}
 -\delta^\nu{}_{[\lambda_1}f_{|\mu|}{}^\rho{}_{\lambda_2\dots\lambda_n]\rho}
\Big)
\nn & &
-g_{\mu[\lambda_1}f_{\lambda_2}{}^\nu{}_{\lambda_3\dots\lambda_n]}
 -\delta^\nu{}_{[\lambda_1}f_{|\mu|\lambda_2\dots\lambda_n]}
\Big\}.
\label{evenD0}
\eea
We have used Bianchi identity $R_{[\mu\nu\lambda]\rho}=0$ to simplify the expressions in the above equations.

\section{A procedure to solve the condition of commutativity}

Let us try to solve the equations \eqref{oddF0}, \eqref{evenF0}, \eqref{oddf0}, \eqref{evenf0},
\eqref{oddD0} and \eqref{evenD0}.
Since \eqref{oddF0} and \eqref{evenF0} contain only 
$F_\mu{}^\nu{}_{\lambda_1\dots\lambda_n}{}^\rho$, let us consider those equations first.
For $n\ge 2$, taking contraction $\lambda_n=\sigma$ in \eqref{evenF0} we obtain
\bea
0 & =  & 2(D-n+1)F_\mu{}^\nu{}_{\lambda_1\dots\lambda_{n-1}}{}^\rho
 -F_\mu{}^\rho{}_{\lambda_1\dots\lambda_{n-1}}{}^\nu
 +(n-1)F_{[\lambda_1}{}^\nu{}_{|\mu|\lambda_2\dots\lambda_{n-1}]}{}^\rho
\nn & &
 -\delta_\mu{}^\rho F_\sigma{}^\nu{}_{\lambda_1\dots\lambda_{n-1}}{}^\sigma
 +(n-1)\Big(
 \delta_\mu{}^\rho F_{[\lambda_1}{}^\nu{}_{\lambda_2\dots\lambda_{n-1}]\sigma}{}^\sigma
 -2F_\mu{}^\nu{}_{[\lambda_1\dots\lambda_{n-2}|\sigma|}{}^\sigma\delta_{\lambda_{n-1}]}{}^\rho
\nn & &
 +F_\mu{}^\rho{}_{[\lambda_1\dots\lambda_{n-2}|\sigma|}{}^\sigma\delta_{\lambda_{n-1}]}{}^\nu
 +F_\mu{}^\sigma{}_{\sigma[\lambda_1\dots\lambda_{n-2}}{}^\rho\delta_{\lambda_{n-1}]}{}^\nu
\Big)
\nn & &
 -n(n+1)\Big(
 F_\mu{}^\sigma{}_{\lambda_1\dots\lambda_{n-1}\sigma}{}^{\nu\rho}
 +F_\mu{}^\rho{}_{\lambda_1\dots\lambda_{n-1}\sigma}{}^{\nu\sigma}
\nn & &
 -F_\sigma{}^\nu{}_{\lambda_1\dots\lambda_{n-1}\mu}{}^{\sigma\rho}
 +\delta_\mu{}^\rho F_\tau{}^\nu{}_{\lambda_1\dots\lambda_{n-1}}{}^\tau{}_\sigma{}^\sigma
\Big).
\label{evenF1}
\eea
Taking contraction $\lambda_n=\nu$ in \eqref{evenF0} we obtain
\bea
0 & =  & (D-n-1)F_\mu{}^\rho{}_{\lambda_1\dots\lambda_{n-1}}{}^\sigma
\nn & &
 +\delta_\mu{}^\rho F_\tau{}^\tau{}_{\lambda_1\dots\lambda_{n-1}}{}^\sigma
\nn & &
-(n-1)\Big(
 \delta_\mu{}^\rho F_{[\lambda_1}{}^\tau{}_{|\tau|\lambda_2\dots\lambda_{n-1}]}{}^\sigma
 -2F_\mu{}^\tau{}_{[\lambda_1\dots\lambda_{n-1}|\tau|}{}^\sigma\delta_{\lambda_{n-1}]}{}^\rho
\Big)
\nn & &
 -n(n+1)\delta_\mu{}^\rho F_{\tau}{}^\phi{}_{\lambda_1\dots\lambda_{n-1}\phi}{}^{\tau\sigma}
\nn & &
 +(\rho\leftrightarrow\sigma).
\label{evenF2}
\eea
Taking contraction $\mu=\sigma$ in \eqref{evenF0} we obtain
\bea
0 & =  & 
 (D-1)F_{[\lambda_n}{}^\nu{}_{\lambda_1\dots\lambda_{n-1}]}{}^\rho
\nn & &
 -2F_\sigma{}^\nu{}_{[\lambda_1\dots\lambda_{n-1}}{}^\sigma\delta_{\lambda_n]}{}^\rho
\nn & &
 +F_\sigma{}^\sigma{}_{[\lambda_1\dots\lambda_{n-1}}{}^\rho\delta_{\lambda_n]}{}^\nu
 +F_\sigma{}^\rho{}_{[\lambda_1\dots\lambda_{n-1}}{}^\sigma\delta_{\lambda_n]}{}^\nu
\nn & &
 +(n+1)\Big(
 F_\sigma{}^\sigma{}_{\lambda_1\dots\lambda_n}{}^{\nu\rho}
 +F_\sigma{}^\rho{}_{\lambda_1\dots\lambda_n}{}^{\nu\sigma}
 -(D+1)F_\sigma{}^\nu{}_{\lambda_1\dots\lambda_n}{}^{\sigma\rho}
\Big).
\label{evenF3}
\eea
The first lines of
\eqref{evenF1}, \eqref{evenF2}, and \eqref{evenF3}
consist of $F_\mu{}^\nu{}_{\lambda_1\dots\lambda_{n-1}}{}^\rho$ with no pair of indices contracted,
and the rest consist of ones with some pairs of indices contracted.
From
\eqref{evenF1}
$+\{(D-n-1)^{-1}\times$\eqref{evenF2} with $\sigma$ replaced by $\nu\}$
$+\{n(D-1)^{-1}\times$\eqref{evenF3} with $\lambda_n$ replaced by $\mu\}$,
we obtain the following expression of $F_\mu{}^\nu{}_{\lambda_1\dots\lambda_{n-1}}{}^\rho$:
\bea
F_\mu{}^\nu{}_{\lambda_1\dots\lambda_{n-1}}{}^\rho
& = & 
\text{(terms proportional to} ~F_\mu{}^\nu{}_{\lambda_1\dots\lambda_{n-1}}{}^\rho ~\text{or}
\nn & &
F_\mu{}^\nu{}_{\lambda_1\dots\lambda_{n+1}}{}^\rho ~\text{with some pairs of indices contracted)}.
\label{evenFf}
\eea
By taking various ways of contractions in this equation we obtain 
relations of $F_\mu{}^\nu{}_{\lambda_1\dots\lambda_{n-1}}{}^\rho$ with some pairs of indices contracted.
Using those relations we can simplify the right hand side of \eqref{evenFf}.
Plugging \eqref{evenFf} into \eqref{evenF0} we may obtain more relations for contracted ones.
After we obtain as much information on the contracted tensors as possible in this way,
some tensors are left undetermined.

Next let us discuss odd $n$ cases.
For $n=1$, taking contraction $\sigma=\rho$ in \eqref{oddF0},
\beq
F_{\mu\nu\lambda} = g_{\mu\nu}F_{\lambda\sigma}{}^\sigma-4F_{\lambda\nu\mu\sigma}{}^\sigma
 +2F_{\lambda\sigma\mu\nu}{}^\sigma + 2F_{\sigma\nu\mu}{}^\sigma{}_\lambda.
\eeq
The right hand side of this equation contains only those with a pair of indices contracted.
However it is not easy to obtain similar expressions for $3\le n\le D-1$ directly.
Therefore
we introduce the following 
dual variables $\wt{F}_\mu{}^\nu{}^{\lambda_{n+1}\dots\lambda_D}{}^\sigma$ for $3\le n\le D-1$:
\beq
F_\mu{}^\nu{}_{\lambda_1\dots\lambda_n}{}^\sigma
 =\frac{1}{(D-n)!}\epsilon_{\lambda_1\dots\lambda_D}\wt{F}_\mu{}^\nu{}^{\lambda_{n+1}\dots\lambda_D}{}^\sigma,
\eeq
or conversely,
\beq
\wt{F}_\mu{}^\nu{}^{\lambda_{n+1}\dots\lambda_D}{}^\sigma
 =\mp\frac{1}{n!}\epsilon^{\lambda_1\dots\lambda_D}F_\mu{}^\nu{}_{\lambda_1\dots\lambda_n}{}^\sigma.
\eeq
Signs in the above equation correspond to the signature of the background metric.
Since all the equations are linear we do not have to care about these signs. 
Then \eqref{oddF0} is rewritten as
\bea
0 & = & (D-n)\Big(
 -2g^{\rho[\lambda_{n+1}}\wt{F}_\mu{}^{|\nu|\lambda_{n+2}\dots\lambda_D]\sigma}
 +g^{\nu[\lambda_{n+1}}\wt{F}_\mu{}^{|\rho|\lambda_{n+2}\dots\lambda_D]\sigma}
\nn & & 
 +\delta_\mu{}^\rho\wt{F}^{[\lambda_{n+1}|\nu|\lambda_{n+2}\dots\lambda_D]\sigma}
\Big)
\nn & & 
 +\frac{1}{n(n+1)}\Big(
 \wt{F}_\mu{}^{|\rho\nu|\lambda_{n+1}\dots\lambda_D\sigma}
 -\delta_\mu{}^\rho\wt{F}_\tau{}^{\nu\tau\lambda_{n+1}\dots\lambda_D\sigma}
\Big)
\nn & &
+(\rho\leftrightarrow\sigma).
\label{oddFt0}
\eea
Taking contraction $\sigma=\lambda_{n+1}$ of this equation, we obtain
\beq
0=-(2n+3)\wt{F}^{\mu\nu\lambda_{n+2}\dots\lambda_D\rho}
 + \wt{F}^{\mu\rho\lambda_{n+2}\dots\lambda_D\nu}
 +(D-n)\wt{F}^{[\mu|\nu|\lambda_{n+2}\dots\lambda_D]\rho}
 +\dots,
\label{oddFt1}
\eeq
and by taking contraction $\nu=\lambda_{n+1}$ and renaming $\sigma$ as $\nu$, we obtain
\beq
0=(n-1)\wt{F}^{\mu\nu\lambda_{n+2}\dots\lambda_D\rho}
 +(n-1)\wt{F}^{\mu\rho\lambda_{n+2}\dots\lambda_D\nu}
 +\dots.
\label{oddFt2}
\eeq
By taking contraction $\sigma=\mu$ and renaming $\lambda_{n+1}$ as $\mu$,
\beq
0=(D-n)(D-1)\wt{F}^{[\mu|\nu|\lambda_{n+2}\dots\lambda_D]\rho}
+\dots.
\label{oddFt3}
\eeq
In \eqref{oddFt1}, \eqref{oddFt2}, and \eqref{oddFt3} ellipses denote 
terms proportional to $\wt{F}_\mu{}^\nu{}_{\lambda_{n+2}\dots\lambda_D}{}^\rho$ or
$\wt{F}_\mu{}^\nu{}_{\lambda_n\dots\lambda_D}{}^\rho$ with some pairs of indices contracted.
From \eqref{oddFt1}$-(n-1)^{-1}\times$\eqref{oddFt2}$-(D-1)^{-1}\times$\eqref{oddFt3},
\bea
\wt{F}_\mu{}^\nu{}_{\lambda_{n+2}\dots\lambda_D}{}^\rho
& = &\text{(terms proportional to} ~\wt{F}_\mu{}^\nu{}_{\lambda_{n+2}\dots\lambda_D}{}^\rho ~\text{or}
\nn & &
 \wt{F}_\mu{}^\nu{}_{\lambda_n\dots\lambda_D}{}^\rho ~\text{with some pairs of indices contracted)}.
\label{oddFtf}
\eea
\eqref{oddFt0} is available also for $n=1$.
In that case \eqref{oddFt2} is not available, but from
\eqref{oddFt1}$-(D-1)^{-1}\times$\eqref{oddFt3} and the same equation with $\rho$ and $\nu$ interchanged
we obtain an expression similar to \eqref{oddFtf}.
For $n=D-1$ \eqref{oddFt3} suffices to obtain \eqref{oddFtf}. 

Next we consider \eqref{oddf0} and \eqref{evenf0}.
In both equations, by taking contraction $\nu=\lambda_n$, we obtain the following for $1\le n\le D$:
\bea
f_\mu{}^\sigma{}_{\lambda_1\dots\lambda_{n-1}}
 & = & (\text{terms proportional to}~
 f_\mu{}^\nu{}_{\lambda_1\dots\lambda_{n-1}},
 f_\mu{}^\nu{}_{\lambda_1\dots\lambda_{n+1}},
\nn & & 
 \nabla_\rho F_\mu{}^\nu{}_{\lambda_1\dots\lambda_{n-1}}{}^\sigma,
 \nabla_\rho F_\mu{}^\nu{}_{\lambda_1\dots\lambda_{n+1}}{}^\sigma,
\nn & & 
 F_\mu{}^\nu{}_{\lambda_1\dots\lambda_{n-2}}{}^\sigma,
 F_\mu{}^\nu{}_{\lambda_1\dots\lambda_n}{}^\sigma,~\text{or}~
 F_\mu{}^\nu{}_{\lambda_1\dots\lambda_{n+2}}{}^\sigma,
\nn & & 
 \text{with some pairs of indices contracted}).
\label{bothf}
\eea
A similar expression for $f_\mu{}^\sigma{}_\lambda$ can be obtained directly from \eqref{evenf0} for $n=0$.
A similar expression for $f_\mu{}^\sigma{}_{\lambda_1\dots\lambda_D}$ is obtained as follows:
\eqref{oddf0} for  $n=D-1$ is expressed as
\bea
0 & = & D(
 2f_\mu{}^\nu{}_{\lambda_1\dots\lambda_{D-1}}{}^\sigma
 -f_\mu{}^\sigma{}_{\lambda_1\dots\lambda_{D-1}}{}^\nu)
\nn & &
+(\text{terms proportional to}~f_\mu{}^\nu{}_{\lambda_1\dots\lambda_{D-2}})
\nn & &
+(\text{terms containing some contracted pairs of indices}).
\label{Dindexedf}
\eea
By plugging \eqref{bothf} with $n=D-1$ into the second line of the right hand side of the above,
it is expressed by terms containing contracted pairs of indices.
Then eliminating $f_\mu{}^\nu{}_{\lambda_1\dots\lambda_{D-1}}{}^\sigma$
from \eqref{Dindexedf} and \eqref{Dindexedf} with $\nu$ and $\sigma$ interchanged,
we obtain an expression of $f_\mu{}^\sigma{}_{\lambda_1\dots\lambda_D}$
by terms containing contracted pairs of indices.

As in the analysis for $F_\mu{}^\nu{}_{\lambda_1\dots\lambda_n}{}^\rho$,
by taking contractions in \eqref{bothf} and similar equations we obtain more 
information on contracted tensors. After obtaining as much information on contracted tensors as possible
we obtain simplified expressions for uncontracted $F_\mu{}^\nu{}_{\lambda_1\dots\lambda_n}{}^\sigma$ and 
$f_\mu{}^\sigma{}_{\lambda_1\dots\lambda_n}$.
Plugging such expressions into \eqref{oddD0} and \eqref{evenD0}, we obtain
more relations for tensors with some pairs of indices contracted, and some tensors are left undetermined.

\section{The solution in $D=4$ case}

Although all equations we are dealing with are linear, 
it is not easy to perform the procedure explained in the previous section for arbitrary $D$ and obtain the most general
solution to \eqref{oddF0}, \eqref{evenF0}, \eqref{oddf0}, \eqref{evenf0}, \eqref{evenD0}, and \eqref{oddD0}.
Therefore in this section we set $D=4$ and solve the equations by brute force calculation by using Mathematica. 
Since it would be very long if we explain the details of the process and it is not illuminating, we only show the result below.

Firstly, the solution to \eqref{oddF0} and \eqref{evenF0} is given by
\bea
F_\mu{}^{\nu\rho} & = &  \delta_\mu{}^\nu k^\rho
 - 2Y_{(1)\mu}{}^{\nu\rho} + 6Y_{(2)\mu}{}^{\nu\rho} + 6Y_{(3)\mu}{}^{\nu\rho}, \\
F_\mu{}^\nu{}_{\lambda_1}{}^\rho & = & c_1(\delta_\mu{}^\nu\delta_{\lambda_1}{}^\rho
 - \delta_\mu{}^\rho\delta_{\lambda_1}{}^\nu)
 -c_2\delta_\mu{}^\nu\delta_{\lambda_1}{}^\rho
 + \delta_\mu{}^\rho K_{\lambda_1}{}^\nu, \\
F_\mu{}^\nu{}_{\lambda_1\lambda_2}{}^\rho & = & 4g^{\nu\sigma}g^{\rho\tau}g_{\mu[\sigma}Y_{(1)\lambda_1\lambda_2\tau]}
 -4\delta_\mu{}^{[\nu}Y_{(2)}{}^{\rho]}{}_{\lambda_1\lambda_2}
 -6g_{\mu[\lambda_1}Y_{(2)\lambda_2]}{}^{\nu\rho}
\nn & &
 -\delta_\mu{}^\nu Y_{(3)}{}^\rho{}_{\lambda_1\lambda_2}
 +2\delta_\mu{}^\rho Y_{(3)}{}^\nu{}_{\lambda_1\lambda_2}
 -2g_{\mu[\lambda_1}Y_{(3)\lambda_2]}{}^{\nu\rho}, \\
F_\mu{}^\nu{}_{\lambda_1\lambda_2\lambda_3}{}^\rho & = &
 c_2g_{\mu[\lambda_1}\delta_{\lambda_2}{}^\nu\delta_{\lambda_3]}{}^\rho
 +\delta_{[\lambda_1}{}^\nu\delta_{\lambda_2}{}^\rho K_{\lambda_3]\mu}
 +\frac{3}{4}\delta_\mu{}^\rho\delta_{[\lambda_1}{}^\nu y_{\lambda_2\lambda_3]}
 +\frac{3}{2}\delta_{[\lambda_1}{}^\nu\delta_{\lambda_2}{}^\rho y_{\lambda_3]\mu}
, \\
F_\mu{}^\nu{}_{\lambda_1\dots\lambda_4}{}^\rho & = & 
 \frac{5}{9}\delta_\mu{}^\nu\delta_{[\lambda_1}{}^\rho (Y_{(1)\lambda_2\lambda_3\lambda_4]}
 -2Y_{(2)\lambda_2\lambda_3\lambda_4]} - 2Y_{(3)\lambda_2\lambda_3\lambda_4]})
\nn & & 
 -\frac{8}{9}\delta_\mu{}^\rho\delta_{[\lambda_1}{}^\nu (Y_{(1)\lambda_2\lambda_3\lambda_4]}
 -2Y_{(2)\lambda_2\lambda_3\lambda_4]} - 2Y_{(3)\lambda_2\lambda_3\lambda_4]})
\nn & & 
 -\frac{5}{3}g_{\mu[\lambda_1}\delta_{\lambda_2}{}^\nu (Y_{(1)\lambda_3\lambda_4]}{}^\rho
 -2Y_{(2)\lambda_3\lambda_4]}{}^\rho - 2Y_{(3)\lambda_3\lambda_4]}{}^\rho)
\nn & & 
 +\frac{5}{3}g_{\mu[\lambda_1}\delta_{\lambda_2}{}^\rho (Y_{(1)\lambda_3\lambda_4]}{}^\nu
 -2Y_{(2)\lambda_3\lambda_4]}{}^\nu - 2Y_{(3)\lambda_3\lambda_4]}{}^\nu),
\eea
where $c_1$ and $c_2$ are scalar functions, $k_\mu$ is a vector,
$y_{\mu\nu}$, $Y_{(1)\lambda_1\lambda_2\lambda_3}$, 
$Y_{(2)\lambda_1\lambda_2\lambda_3}$, and $Y_{(3)\lambda_1\lambda_2\lambda_3}$ are antisymmetric tensors,
and $K_{\mu\nu}$ is a symmetric tensor satisfying
\beq
K_\mu{}^\mu=0.
\eeq
At this stage these tensors are independent and have no more restriction.

Next, the solution to \eqref{oddf0} and \eqref{evenf0} is given by
\bea
f_\mu{}^\nu & = & \Big(c_0+\frac{3}{2}mc_1\Big)\delta_\mu{}^\nu+\nabla_\mu k^\nu, \\
f_\mu{}^\nu{}_{\lambda_1} & = & 0,\\
f_\mu{}^\nu{}_{\lambda_1\lambda_2} & = &
 \frac{1}{4}c_1m(g_{\mu\lambda_1}\delta_{\lambda_2}{}^\nu
 -g_{\mu\lambda_2}\delta_{\lambda_1}{}^\nu)
 +\frac{1}{4}\delta_\mu{}^\nu\nabla_{\lambda_1}k_{\lambda_2}
\nn & & 
 +\frac{1}{2}g^{\nu\rho}\nabla_{[\mu}Y_{(1)\rho\lambda_1\lambda_2]}
 -g^{\nu\rho}\nabla_{[\mu}Y_{(2)\rho\lambda_1\lambda_2]}
\nn & & 
 +\frac{1}{16}\delta_\mu{}^\nu\nabla_\rho Y_{(1)}{}^\rho{}_{\lambda_1\lambda_2}
 -\frac{1}{8}g_{[\lambda_1}{}^\nu\nabla_{|\rho|}Y_{(1)}{}^\rho{}_{\lambda_2]\mu}, \\
f_\mu{}^\nu{}_{\lambda_1\lambda_2\lambda_3} & = & 0,\\
f_\mu{}^\nu{}_{\lambda_1\dots\lambda_4} & = & 
 \frac{7}{144}\delta_\mu{}^\nu \hat{Y}_{(1)\lambda_1\dots\lambda_4}
 +\frac{1}{9}g_{\mu[\lambda_1}\hat{Y}_{(1)\lambda_2\lambda_3\lambda_4]}{}^\nu
\nn & &
 -\frac{5}{72}\delta_\mu{}^\nu \hat{Y}_{(2)\lambda_1\dots\lambda_4}
 -\frac{4}{9}g_{\mu[\lambda_1}\hat{Y}_{(2)\lambda_2\lambda_3\lambda_4]}{}^\nu
 +\frac{5}{16}g_{\mu[\lambda_1}\delta_{\lambda_2}{}^\nu\nabla_{|\rho|}Y_{(1)}{}^\rho{}_{\lambda_3\lambda_4]},
\eea
where $c_0$ is a scalar function, and
\beq
\hat{Y}_{(1)\lambda_1\dots\lambda_4}\equiv\nabla_{[\lambda_1}Y_{(1)\lambda_2\lambda_3\lambda_4]},\quad
\hat{Y}_{(2)\lambda_1\dots\lambda_4}\equiv\nabla_{[\lambda_1}Y_{(2)\lambda_2\lambda_3\lambda_4]}.
\eeq
It turns out that $c_1$ and $c_2$ are constants,
$k_\mu$ is a Killing vector, and 
\beq
mK_{\mu\nu}=0,\quad \nabla_\mu K_{\nu\rho}=0,
\eeq
\beq
m\Big(Y_{(2)\lambda_1\lambda_2\lambda_3}-\frac{1}{4}Y_{(1)\lambda_1\lambda_2\lambda_3}\Big)=0,\quad
m\Big(Y_{(3)\lambda_1\lambda_2\lambda_3}-\frac{1}{4}Y_{(1)\lambda_1\lambda_2\lambda_3}\Big)=0,
\eeq
\beq
\nabla_\rho Y_{(1)}{}^\rho{}_{\lambda_1\lambda_2}=6my_{\lambda_1\lambda_2},\quad
\nabla_\mu y_{\nu\rho}=0,
\label{abouty}
\eeq
\bea
\nabla_\mu Y_{(1)\lambda_1\dots\lambda_3} & = &
 \hat{Y}_{(1)\mu\lambda_1\dots\lambda_3}
 +\frac{3}{2}g_{\mu[\lambda_1}\nabla_{|\rho|}Y_{(1)}{}^\rho{}_{\lambda_2\lambda_3]},
\label{aboutY1} \\
\nabla_\mu Y_{(2)\lambda_1\dots\lambda_3} & = &
 \hat{Y}_{(2)\mu\lambda_1\dots\lambda_3}
 +\frac{3}{8}g_{\mu[\lambda_1}\nabla_{|\rho|}Y_{(1)}{}^\rho{}_{\lambda_2\lambda_3]}, \\
\nabla_\mu Y_{(3)\lambda_1\dots\lambda_3} & = &
 \frac{1}{2}\hat{Y}_{(1)\mu\lambda_1\dots\lambda_3} - \hat{Y}_{(2)\mu\lambda_1\dots\lambda_3}.
\label{aboutY3}
\eea

Next, using the above solutions we can give the solution to \eqref{oddD0} and \eqref{evenD0}.
It turns out that $c_0$ is a constant, and
\beq
m\nabla_\rho Y_{(1)}{}^\rho{}_{\lambda_2\lambda_2}=0.
\eeq
From this and \eqref{abouty}, 
\beq
\nabla_\rho Y_{(1)}{}^\rho{}_{\lambda_2\lambda_2}=0, \quad my_{\lambda_1\lambda_2}=0.
\eeq
Therefore from \eqref{aboutY1}-\eqref{aboutY3}, it turns out that
$Y_{(1)\lambda_1\dots\lambda_3}$, $Y_{(2)\lambda_1\dots\lambda_3}$, 
and $Y_{(3)\lambda_1\dots\lambda_3}$ are Killing-Yano tensors.
Instead of these tensors, we introduce $Y_{\lambda_1\lambda_2\lambda_3}$, 
$U_{\lambda_1\lambda_2\lambda_3}$ and $V_{\lambda_1\lambda_2\lambda_3}$ defined as follows:
\bea
Y_{\lambda_1\lambda_2\lambda_3} & \equiv & Y_{(1)\lambda_1\lambda_2\lambda_3}, \\
U_{\lambda_1\lambda_2\lambda_3} & \equiv & Y_{(2)\lambda_1\lambda_2\lambda_3}
 -\frac{1}{4}Y_{\lambda_1\lambda_2\lambda_3}, \\
V_{\lambda_1\lambda_2\lambda_3} & \equiv & Y_{(3)\lambda_1\lambda_2\lambda_3}
 -\frac{1}{4}Y_{\lambda_1\lambda_2\lambda_3}+U_{\lambda_1\lambda_2\lambda_3}.
\eea
These are also Killing-Yano tensors, and 
\beq
mU_{\lambda_1\lambda_2\lambda_3}=0, \quad mV_{\lambda_1\lambda_2\lambda_3}=0.
\eeq
$K_{\mu\nu}, y_{\mu\nu}, Y_{\lambda_1\lambda_2\lambda_3},
U_{\lambda_1\lambda_2\lambda_3}$, and $V_{\lambda_1\lambda_2\lambda_3}$ obey
\bea
R_{\mu[\lambda_1}K_{\lambda_2]\nu} + R_{\nu[\lambda_1}K_{\lambda_2]\mu} & = &
 c_2(g_{\mu[\lambda_1}R_{\lambda_2]\nu} + g_{\nu[\lambda_1}R_{\lambda_2]\mu}),
\label{aboutc2} \\
R_{\rho\sigma\mu\nu}y^{\rho\sigma} & = & Ry_{\mu\nu} , \\
0 & = & R_{\mu[\lambda_1}y_{\lambda_2\lambda_3]},
\eea
\bea
R_{\rho\lambda_1}Y^\rho{}_{\lambda_2\lambda_3} & = &
 R_{\rho[\lambda_1}Y^\rho{}_{\lambda_2\lambda_3]}, \\
R_{\rho\sigma\lambda_1\lambda_2}U^{\rho\sigma}{}_{\lambda_3} & = & 0, \\
R_{\rho\lambda_1}V^\rho{}_{\lambda_2\lambda_3} & = & 0, \\
RV_{\lambda_1\lambda_2\lambda_3} & = & 0.
\eea
Then we obtain the full solution: 
\bea
F_\mu{}^{\nu\rho} & = &  \delta_\mu{}^\nu k^\rho
 + Y_{\mu}{}^{\nu\rho} + 6V_{\mu}{}^{\nu\rho}, \\
F_\mu{}^\nu{}_{\lambda_1}{}^\rho & = & c_1(\delta_\mu{}^\nu\delta_{\lambda_1}{}^\rho
 - \delta_\mu{}^\rho\delta_{\lambda_1}{}^\nu)
 -c_2\delta_\mu{}^\nu\delta_{\lambda_1}{}^\rho
 + \delta_\mu{}^\rho K_{\lambda_1}{}^\nu, \\
F_\mu{}^\nu{}_{\lambda_1\lambda_2}{}^\rho & = & \frac{1}{4}\delta_\mu{}^\nu Y_{\lambda_1\lambda_2}{}^\rho
 -\delta_\mu{}^\nu U_{\lambda_1\lambda_2}{}^\rho
 -4g_{\mu[\lambda_1}U_{\lambda_2]}{}^{\nu\rho}
\nn & &
 +2\delta_\mu{}^\rho V_{\lambda_1\lambda_2}{}^\nu
 -\delta_\mu{}^\nu V_{\lambda_1\lambda_2}{}^\rho
-2g_{\mu[\lambda_1}V_{\lambda_2]}{}^{\nu\rho}, \\
F_\mu{}^\nu{}_{\lambda_1\lambda_2\lambda_3}{}^\rho & = &
 c_2g_{\mu[\lambda_1}\delta_{\lambda_2}{}^\nu\delta_{\lambda_3]}{}^\rho
 +\delta_{[\lambda_1}{}^\nu\delta_{\lambda_2}{}^\rho K_{\lambda_3]\mu}
 +\frac{3}{4}\delta_\mu{}^\rho\delta_{[\lambda_1}{}^\nu y_{\lambda_2\lambda_3]}
 +\frac{3}{2}\delta_{[\lambda_1}{}^\nu\delta_{\lambda_2}{}^\rho y_{\lambda_3]\mu}
, \\
F_\mu{}^\nu{}_{\lambda_1\dots\lambda_4}{}^\rho & = & 
 -\frac{10}{9}\delta_\mu{}^\nu\delta_{[\lambda_1}{}^\rho V_{\lambda_2\lambda_3\lambda_4]}
 +\frac{16}{9}\delta_\mu{}^\rho\delta_{[\lambda_1}{}^\nu V_{\lambda_2\lambda_3\lambda_4]}
 \nn & & 
 +\frac{10}{3}g_{\mu[\lambda_1}\delta_{\lambda_2}{}^\nu V_{\lambda_3\lambda_4]}{}^\rho
 -\frac{10}{3}g_{\mu[\lambda_1}\delta_{\lambda_2}{}^\rho V_{\lambda_3\lambda_4]}{}^\nu,
\eea
\bea
f_\mu{}^\nu & = &\Big(c_0+\frac{3}{2}mc_1\Big)\delta_\mu{}^\nu+\nabla_\mu k^\nu, \\
f_\mu{}^\nu{}_{\lambda_1} & = & 0,\\
f_\mu{}^\nu{}_{\lambda_1\lambda_2} & = &
 \frac{1}{4}c_1m(g_{\mu\lambda_1}\delta_{\lambda_2}{}^\nu
 -g_{\mu\lambda_2}\delta_{\lambda_1}{}^\nu)
\nn & &
 +\frac{1}{4}\delta_\mu{}^\nu\nabla_{[\lambda_1}k_{\lambda_2]}
 +\frac{1}{4}g^{\nu\rho}\nabla_{[\mu}Y_{\rho\lambda_1\lambda_2]}
 -g^{\nu\rho}\nabla_{[\mu}U_{\rho\lambda_1\lambda_2]}, \\
f_\mu{}^\nu{}_{\lambda_1\lambda_2\lambda_3} & = & 0,\\
f_\mu{}^\nu{}_{\lambda_1\dots\lambda_4} & = & 
 \frac{1}{32}\delta_\mu{}^\nu \nabla_{[\lambda_1}Y_{\lambda_2\lambda_3\lambda_4]}
 -\frac{5}{72}\delta_\mu{}^\nu \nabla_{[\lambda_1}U_{\lambda_2\lambda_3\lambda_4]}
 -\frac{4}{9}g_{\mu[\lambda_1}\nabla_{\lambda_2}U_{\lambda_3\lambda_4]}{}^\nu.
\eea
$c_0$ gives part proportional to the unit matrix, and $c_1$ gives part proportional to $\hat{M}_\mu{}^\nu$.
Since these trivially commute with $\hat{M}_\mu{}^\nu$, we set $c_0=c_1=0$ in the following.

If the mass parameter $m$ is nonzero, then
\beq
K_{\mu\nu} =  0, \quad
y_{\mu\nu} = 0, \quad
U_{\lambda_1\lambda_2\lambda_3} = 0, \quad
V_{\lambda_1\lambda_2\lambda_3} = 0,
\eeq
and we are left with a constant $c_2$, a Killing vector $k_\mu$ and
a Killing-Yano tensor $Y_{\lambda_1\lambda_2\lambda_3}$.

Taking contraction $\nu=\lambda_2$ in \eqref{aboutc2} and symmetrizing $\mu$ and $\lambda_1$ we obtain
\beq
RK_{\mu\nu}=4c_2\Big(R_{\mu\nu}-\frac{1}{4}Rg_{\mu\nu}\Big).
\eeq
Therefore if $K_{\mu\nu}=0$, $c_2$ also vanishes unless
\beq
R_{\mu\nu}-\frac{1}{4}Rg_{\mu\nu}=0.
\label{c2not0}
\eeq

\section{Operators commuting with 
$M_\mu{}^\nu$ for arbitrary $D$}

In this section we make an attempt to give operators commuting with $M_\mu{}^\nu$ for arbitrary $D$,
not necessarily even, extending the result for $\hat{M}_\mu{}^\nu$ in the previous section.

We begin with the result for massive case in the previous section, 
which is also a special case of the general form for massless case:
\bea
O_{\mu\nu} & = & c_2(\Gamma_{\mu\nu}{}^\rho-g_{\mu\nu}\Gamma^\rho)\mathcal{D}_\rho
\nn & &
 +g_{\mu\nu}k^\rho\mathcal{D}_\rho + \nabla_{[\mu}k_{\nu]}
 +\frac{1}{4}g_{\mu\nu}\Gamma^{\lambda_1\lambda_2}\nabla_{\lambda_1}k_{\lambda_2}
\nn & &
 +Y_{\mu\nu}{}^\rho\mathcal{D}_\rho
 +\frac{1}{4}g_{\mu\nu}\Gamma^{\lambda_1\lambda_2}Y_{\lambda_1\lambda_2}{}^\rho\mathcal{D}_\rho
\nn & &
 +\frac{1}{4}\Gamma^{\lambda_1\lambda_2}\nabla_{[\mu}Y_{\nu\lambda_1\lambda_2]}
 +\frac{1}{32}g_{\mu\nu}\Gamma^{\lambda_1\dots\lambda_4}
  \nabla_{\lambda_1}Y_{\lambda_2\dots\lambda_4}.
\label{D4result}
\eea
It is not difficult to show that \eqref{D4result} satisfies \eqref{mhattom}, 
and therefore it commutes not only with $\hat{M}_\mu{}^\nu$, but also with $M_\mu{}^\nu$.
Let us try to extend the part proportional to $Y_{\lambda_1\lambda_2\lambda_3}$ to 
arbitrary $D$.
For $n$ even and $0\le n\le D-3$, we consider the following operator constructed of a rank $n+3$ Killing-Yano tensor
$Y_{\lambda_1\dots\lambda_{n+3}}$: 
\bea
O^{(n+3)}_{\mu\nu} & \equiv & (n+2)\Gamma^{\lambda_1\dots\lambda_n}Y_{\mu\nu\lambda_1\dots\lambda_n}{}^\rho\mathcal{D}_\rho
 +\frac{1}{2}g_{\mu\nu}\Gamma^{\lambda_1\dots\lambda_{n+2}}Y_{\lambda_1\dots\lambda_{n+2}}{}^\rho\mathcal{D}_\rho
\nn & &
 +\frac{1}{2}\Gamma^{\lambda_1\dots\lambda_{n+2}}\nabla_{[\mu}Y_{\nu\lambda_1\dots\lambda_{n+2}]}
 +\frac{1}{4(n+4)}g_{\mu\nu}\Gamma^{\lambda_1\dots\lambda_{n+4}}
  \nabla_{\lambda_1}Y_{\lambda_2\dots\lambda_{n+4}}.
\label{KYop}
\eea
If we set $n=-2$ in the above, we obtain an extension of the part proportional to $k_\mu$ in \eqref{D4result}:
\bea
O^{(1)}_{\mu\nu} & = & 
 \frac{1}{2}g_{\mu\nu}Y^\rho\mathcal{D}_\rho + \frac{1}{2}\nabla_{[\mu}Y_{\nu]}
 +\frac{1}{8}g_{\mu\nu}\Gamma^{\lambda_1\lambda_2}\nabla_{\lambda_1}Y_{\lambda_2},
\eea
and it is not difficult to confirm that it commutes with $M_\mu{}^\nu$ for arbitrary $D$
if $Y_\mu$ is a Killing vector.

Furthermore, for $0\le n\le D-3$, we can show the following facts
by straightforward calculation:
\begin{itemize}
\item 
$O^{(n+3)}{}_\mu{}^\nu$ commutes with the mass term in $M_\mu{}^\nu$ 
as long as $Y_{\lambda_1\dots\lambda_{n+3}}$ is an antisymmetric tensor.
\item $[O^{(n+3)},M]_\mu{}^\nu$ has no second order derivative operator as long as
 $Y_{\lambda_1\dots\lambda_{n+3}}$ is an antisymmetric tensor.
\item $[O^{(n+3)},M]_\mu{}^\nu$ has no first order derivative operator
 if $Y_{\lambda_1\dots\lambda_{n+3}}$ is a Killing-Yano tensor.
\item $[O^{(n+3)},M]_\mu{}^\nu=0$ gives the following additional condition:
\beq
R_{\rho\lambda_1}Y^\rho{}_{\lambda_2\dots\lambda_{n+3}} = R_{\rho[\lambda_1}Y^\rho{}_{\lambda_2\dots\lambda_{n+3}]}.
\label{condY1}
\eeq
\end{itemize}
This is consistent with the result in the previous section. 
To show these facts we used \eqref{ddKY} and \eqref{RKY}.

Hodge duals of Killing-Yano tensors are CCKY tensors, and
if $D$ is odd, we can also dualize gamma matrices:
\beq
\Gamma^{\mu_1\dots\mu_p}
 =\pm\frac{1}{(D-p)!}(-1)^{(D-p)(D-p-1)/2}\epsilon^{\mu_1\dots\mu_D}\Gamma_{\mu_{p+1}\dots\mu_D},
\eeq
where the overall sign in the above depends on the signature of the background metric.
Using this we can obtain the dual expression of $O^{(n+3)}_{\mu\nu}$.
For odd $D$ it does not give new commuting operator,
but if we extend such an expression to even $D$, it is not related to \eqref{KYop} and give new operator,
which is denoted by $\wt{O}^{(n)}_{\mu\nu}$:
\bea
\wt{O}^{(n)}{}_\mu{}^\nu & = & \Big\{\Gamma_\mu{}^{\nu\rho\lambda_1\dots\lambda_n}
 -\frac{1}{2}(D-n-2)\delta_\mu{}^\nu\Gamma^{\rho\lambda_1\dots\lambda_n}\Big\}
 C_{\lambda_1\dots\lambda_n}\mathcal{D}_\rho
\nn & &
 +\frac{n}{2}\frac{D-n-2}{D-n+1}\Big\{\Gamma_\mu{}^{\nu\lambda_1\dots\lambda_{n-1}}
 -\frac{1}{2}(D-n)\delta_\mu{}^\nu\Gamma^{\lambda_1\dots\lambda_{n-1}}
 \Big\}\nabla^\rho C_{\rho\lambda_1\dots\lambda_{n-1}},
\label{CCKYop}
\eea
where $n$ is even and $0\le n\le D-1$, and $C_{\lambda_1\dots\lambda_n}$ is CCKY tensors.
If $D$ is odd, $C_{\lambda_1\dots\lambda_n}$ is proportional to
the Hodge dual of $Y_{\lambda_1\dots\lambda_{D-n}}$. For even $D$, the range of $n$ can be restricted to
$0\le n\le D-4$ because $D-1$ and $D-3$ is odd, and $\wt{O}^{(D-2)}{}_\mu{}^\nu$ vanishes.
For arbitrary $D$ we can show the following facts:
\begin{itemize}
\item 
$\wt{O}^{(n)}{}_\mu{}^\nu$ commutes with the mass term in $M_\mu{}^\nu$ 
as long as $C_{\lambda_1\dots\lambda_n}$ is an antisymmetric tensor.
\item $[\wt{O}^{(n)},M]_\mu{}^\nu$ has no second order derivative operator as long as
 $C_{\lambda_1\dots\lambda_n}$ is an antisymmetric tensor.
\item $[\wt{O}^{(n)},M]_\mu{}^\nu$ has no first order derivative operator
 if $C_{\lambda_1\dots\lambda_n}$ is a CCKY tensor.
\item $[\wt{O}^{(n)},M]_\mu{}^\nu=0$ gives the following additional condition:
\beq
R_{\mu[\lambda_1}C_{\lambda_2\dots\lambda_{n+1}]}
 = \frac{1}{D-n}g_{\mu[\lambda_1}
 (RC_{\lambda_2\dots\lambda_{n+1}]}-nR_{|\rho|\lambda_2}C^\rho{}_{\lambda_3\dots\lambda_{n+1}]}),
\label{condY2}
\eeq
which is equivalent to \eqref{condY1} if $D$ is odd.
\end{itemize}
To show these facts we used \eqref{ddCCKY} and \eqref{RCCKY}.
Note that both \eqref{condY1} and \eqref{condY2} are satisfied if the background geometry
obeys vacuum Einstein equation, with or without cosmological term.
For $n=0$, $C=C_{\lambda_1\dots\lambda_n}$ is a constant.
Then for $D=4$, $\wt{O}^{(0)}{}_\mu{}^\nu$ and \eqref{condY2} reproduce the part proportional to 
$c_2$ in \eqref{D4result} and \eqref{c2not0}.

Thus we have found that 
\beq
O_\mu{}^\nu = \sum_{n=0, \text{even}}^{D-2}O^{(n+1)}{}_\mu{}^\nu
 +\sum_{n=0, \text{even}}^{D-4}\wt{O}^{(n)}{}_\mu{}^\nu,
\label{opDeven}
\eeq
for even $D$, and
\beq
O_\mu{}^\nu = \sum_{n=0, \text{even}}^{D-1}O^{(n+1)}{}_\mu{}^\nu,
\label{opDodd}
\eeq
for odd $D$ commute with $M_\mu{}^\nu$ if the conditions \eqref{condY1} and \eqref{condY2} are satisfied.

Note that $O^{(n+1)}{}_\mu{}^\nu$ and $\wt{O}^{(n)}{}_\mu{}^\nu$ satisfy
\bea
\Gamma^\mu O^{(n+1)}{}_\mu{}^\nu & = & \frac{1}{2}K^{(n+1)} \Gamma^\nu,
\label{oggk1} \\
\Gamma^\mu \wt{O}^{(n)}{}_\mu{}^\nu & = & -\frac{1}{2}(D-n-2)\wt{K}^{(n)} \Gamma^\nu,
\label{oggk2}
\eea
and
\bea
O^{(n+1)}{}_\mu{}^\nu \Gamma_\nu & = & \frac{1}{2}\Gamma_\mu K^{(n+1)}, \\
\wt{O}^{(n)}{}_\mu{}^\nu \Gamma_\nu & = & -\frac{1}{2}(D-n-2)\Gamma_\mu \wt{K}^{(n)},
\label{oggk3}
\eea
where
\bea
K^{(n+1)} & = & \Gamma^{\lambda_1\dots\lambda_n}Y_{\lambda_1\dots\lambda_n}{}^\rho\mathcal{D}_\rho
 +\frac{1}{2(n+2)}\Gamma^{\lambda_1\dots\lambda_{n+2}}
  \nabla_{\lambda_1}Y_{\lambda_2\dots\lambda_{n+2}}, \\
\wt{K}^{(n)} & = & \Gamma^{\rho\lambda_1\dots\lambda_n}C_{\lambda_1\dots\lambda_n}\mathcal{D}_\rho
 +\frac{1}{2}\frac{n(D-n)}{D-n+1}\Gamma^{\lambda_1\dots\lambda_{n-1}}
  \nabla^\rho C_{\rho\lambda_1\dots\lambda_{n-1}},
\eea
are the operators commuting with the Dirac operator\cite{bc96,ckk11}. 
(To be precise, $\mathcal{D}_\rho$ is supposed to act on Rarita-Schwinger fields.)
These relations hold as long as $Y_{\lambda_1\dots\lambda_{n+1}}$ and 
$C_{\lambda_1\dots\lambda_n}$ are antisymmetric tensors.

\section{Discussion}

In section 2, 3 and 4 we have considered $\hat{M}_\mu{}^\nu$ instead of $M_\mu{}^\nu$ to make calculations easier.
However if we use the background field equation from the beginning we can simplify it more 
as is often done in the literature:
From the equations given by acting $\mathcal{D}_\mu$ on \eqref{rseq}
and by multiplying $m\Gamma_\mu$ on \eqref{rseq}, we obtain
\beq
\Big(R_\mu{}^\nu-\frac{1}{2}\delta_\mu{}^\nu R
 -2\frac{D-1}{D-2}m^2\delta_\mu{}^\nu\Big)\Gamma^\mu\psi_\nu=0.
\eeq
If the background metric satisfies vacuum Einstein equation
\beq
R_{\mu\nu}-\frac{1}{2}g_{\mu\nu}R+\Lambda g_{\mu\nu}=0,
\eeq
then
\beq
\Big(\Lambda+2\frac{D-1}{D-2}m^2\Big)\Gamma^\mu\psi_\mu=0.
\eeq
Unless $\Lambda=-2\frac{D-1}{D-2}m^2$, we obtain
\beq
\Gamma^\mu\psi_\mu=0,
\eeq
and using this \eqref{rseq} can be simplified to
\beq
(\Gamma^\mu\mathcal{D}_\mu+m)\psi_\nu=0.
\eeq
Conversely it is easy to show that \eqref{rseq} follows the above two equations. 
If $\Lambda=-2\frac{D-1}{D-2}m^2$, \eqref{rseq} has a gauge symmetry 
$\delta\psi_\mu=\mathcal{D}_\mu\epsilon+\frac{m}{D-2}\Gamma_\mu\epsilon$,
and by imposing gauge fixing condition $\Gamma^\mu\psi_\mu=0$ \eqref{rseq} is reduced to 
$(\Gamma^\mu\mathcal{D}_\mu+m)\psi_\nu=0$.
Thus \eqref{rseq} is reduced to `Dirac-like' equation $(\Gamma^\mu\mathcal{D}_\mu+m)\psi_\nu=0$.
Although this is not ordinary Dirac equation
because the covariant derivative acts on vector-spinors differently from spinors,
the difference is not so significant, as the first term in \eqref{comcd}.
This `Dirac-like' equation simplifies the analysis further.
However, in general, operators commuting with this `Dirac-like' operator does not directly give quantum numbers of
the solutions due to the condition $\Gamma^\mu\psi_\mu=0$.
Furthermore if we have more nontrivial background fields such as electromagnetic field, then
Einstein equation is modified and the above argument does not hold.

In section 5, for arbitrary $D$ we have constructed first order operators which commute with
$M_\mu{}^\nu$, and are constructed of odd rank Killing-Yano tensors and even rank CCKY tensors.
Although we have additional conditions \eqref{condY1} and \eqref{condY2}, 
this situation is similar to the case of operators commuting with the Dirac operator.
Indeed our operators $O^{(n+1)}{}_\mu{}^\nu$ and $\wt{O}^{(n)}{}_\mu{}^\nu$ are related to 
$K^{(n+1)}$ and $\wt{K}^{(n)}$ by \eqref{oggk1}-\eqref{oggk3}.
Especially \eqref{oggk1} and \eqref{oggk2} means that $O^{(n+1)}{}_\mu{}^\nu$ and $\wt{O}^{(n)}{}_\mu{}^\nu$
preserve the condition $\Gamma^\mu\psi_\mu=0$. Therefore $O^{(n+1)}{}_\mu{}^\nu$ and $\wt{O}^{(n)}{}_\mu{}^\nu$
are extensions of $K^{(n+1)}$ and $\wt{K}^{(n)}$ to the space of Rarita-Schwinger fields.
In the case of the Dirac operator it is proven that there are no other commuting operators than 
$K^{(n+1)}$ and $\wt{K}^{(n)}$ \cite{ckk11}.
It is desirable to clarify if there is any other operator commuting with $M_\mu{}^\nu$. 
Moreover if we have two or more odd rank Killing-Yano, or even rank CCKY tensors,
it is necessary to know about the commutativity between the operators constructed of different
Killing-Yano or CCKY tensors, or other operators commuting with $M_\mu{}^\nu$.

We first considered $\hat{M}_\mu{}^\nu$, 
and in general, operators commuting with $\hat{M}_\mu{}^\nu$ are symmetry operators
for $M_\mu{}^\nu$. However a symmetry operator for $\hat{M}_\mu{}^\nu$ is also
a symmetry operator for  $M_\mu{}^\nu$ and vice versa.
Therefore it may be helpful to use $\hat{M}_\mu{}^\nu$ to 
give general form of symmetry operators for $M_\mu{}^\nu$.

\vs{.5cm}
\noindent
{\large\bf Acknowledgments}\\[.2cm]
I would like to thank T.~Suyama for correspondence.

\renewcommand{\theequation}{\Alph{section}.\arabic{equation}}
\appendix
\addcontentsline{toc}{section}{Appendix}
\vs{.5cm}
\noindent
{\Large\bf Appendix}
\section{Killing-Yano and closed conformal Killing-Yano tensors}
\label{appa}
\setcounter{equation}{0}

A Killing-Yano tensor $Y_{\lambda_1\dots\lambda_n}$ is defined as an antisymmetric tensor obeying
\beq
\nabla_\mu Y_{\lambda_1\dots\lambda_n} = \nabla_{[\mu}Y_{\lambda_1\dots\lambda_n]}.
\eeq
From this equation we can show the following:
\beq
\nabla_\mu\nabla_\nu Y_{\lambda_1\dots\lambda_n}
 = -\frac{(n+1)}{2}R_{\mu\rho[\nu\lambda_1}Y^\rho{}_{\lambda_2\dots\lambda_n]},
\label{ddKY}
\eeq
\beq
0=R_{\mu\sigma_1[\nu}{}^\rho Y_{|\rho|\sigma_2\lambda_1\dots\lambda_{n-2}]}
+R_{\nu\sigma_1[\mu}{}^\rho Y_{|\rho|\sigma_2\lambda_1\dots\lambda_{n-2}]}
+(\sigma_1\leftrightarrow\sigma_2).
\label{RKY}
\eeq
The following relations derived from the above are also useful.
\beq
R_{\rho\mu}Y^\rho{}_{\nu\lambda_1\dots\lambda_{n-2}}+(\mu\leftrightarrow\nu)
=-\frac{1}{2}(n-2)(R_{\rho\sigma\mu[\lambda_1}Y^{\rho\sigma}{}_{|\nu|\lambda_2\dots\lambda_{n-2}]}
 +(\mu\leftrightarrow\nu)),
\eeq
\beq
R_{\mu\nu\rho[\lambda_1}Y^\rho{}_{\lambda_2\dots\lambda_n]}
=\frac{1}{2}R_{[\lambda_1\lambda_2|\mu\rho}Y^\rho{}_{\nu|\lambda_3\dots\lambda_n]}
 -\frac{1}{2}R_{[\lambda_1\lambda_2|\nu\rho}Y^\rho{}_{\mu|\lambda_3\dots\lambda_n]}.
\eeq

A conformal Killing-Yano (CKY) tensor $C_{\lambda_1\dots\lambda_n}$ is defined as an antisymmetric tensor obeying
\beq
\nabla_\mu C_{\lambda_1\dots\lambda_n}
 = \nabla_{[\mu}C_{\lambda_1\dots\lambda_n]}
 +\frac{n}{D-n+1}g_{\mu[\lambda_1}\nabla_{|\rho|}C^\rho{}_{\lambda_2\dots\lambda_n]}.
\eeq
If $C_{\lambda_1\dots\lambda_n}$ is a closed conformal Killing-Yano (CCKY) tensor,
the first term of the right hand side of the above vanishes.
The Hodge dual of a Killing-Yano tensor is a CCKY tensor.

A CCKY tensor satisfies the following:
\beq
\nabla_\mu\nabla_\nu C_{\lambda_1\dots\lambda_n}
 = \frac{n}{D-n}g_{\nu[\lambda_1}\Big\{-R_{|\mu\rho|}C^\rho{}_{\lambda_2\dots\lambda_n]}
 +\frac{1}{2}(n-1)R_{|\rho\sigma\mu|\lambda_2}C^{\rho\sigma}{}_{\lambda_3\dots\lambda_n]}\Big\},
\label{ddCCKY}
\eeq
\beq
R_{\mu\nu\rho[\lambda_1}C^\rho{}_{\lambda_2\dots\lambda_n]} =
\frac{1}{D-n}g_{\mu[\lambda_1}\Big(-R_{|\nu\rho|}C^\rho{}_{\lambda_2\dots\lambda_n]}
 +\frac{1}{2}(n-1)R_{|\lambda\rho\nu|\lambda_2}C^{\lambda\rho}{}_{\lambda_3\dots\lambda_n]}\Big)
-(\mu\leftrightarrow\nu).
\label{RCCKY}
\eeq

\section{Results in a coordinate-free notation}
\label{appb}
\setcounter{equation}{0}

For readers' convenience we give our results \eqref{KYop}, \eqref{condY1}, \eqref{CCKYop}, and \eqref{condY2}
in the notation used in \cite{bc96,bk04,ckk11}, which is often used for analyses related to the Dirac operator.
First let $Y$ be an odd inhomogeneous Killing-Yano form, and $C$ be an even inhomogeneous CCKY form:
\beq
Y=\sum_{n: \text{even}} \frac{1}{(n+3)!}Y_{\mu_1\mu_2\dots\mu_{n+3}}
 dx^{\mu_1}\wedge dx^{\mu_2}\wedge\dots\wedge dx^{\mu_{n+3}},
\eeq
\beq
C=\sum_{n: \text{even}} \frac{1}{n!}C_{\mu_1\mu_2\dots\mu_n}
 dx^{\mu_1}\wedge dx^{\mu_2}\wedge\dots\wedge dx^{\mu_n}.
\eeq
Then \eqref{KYop} and \eqref{CCKYop} are expressed as follows:
\beq
\sum_{n: \text{even}} \frac{1}{n!}O^{(n+3) ab} = 
 \Big[-X^a\hook X^b\hook + \frac{1}{2}g^{ab}(\pi-1)\Big]
 \Big[\pi X^c\hook Y\nabla_c + \frac{(\pi-2)(\pi-3)}{2\pi}dY \Big],
\eeq
and
\beq
\sum_{n: \text{even}}\frac{1}{n!}\wt{O}^{(n) ab} =  
 \Big[e^a\wedge e^b - \frac{1}{2}g^{ab}(D-\pi-1)\Big]\wedge
 \Big[e^c\wedge C \nabla_c - \frac{D-\pi-3}{2(D-\pi)}\delta C \Big],
\eeq
where we insert an additional factor $\frac{1}{n!}$ which is absent in 
\eqref{opDeven} and \eqref{opDodd} to obtain simple expressions.
The condition \eqref{condY1} and \eqref{condY2} are expressed as follows:
\beq
R_{ba}X^b\hook Y = \frac{1}{\pi +1}X_a\hook[R_{bc} e^c\wedge(X^b\hook Y)],
\eeq
\beq
R_{ab} e^b\wedge C = \frac{1}{D-\pi+1} e_a\wedge[RC-R_{bc} e^c\wedge (X^b\hook C)].
\eeq

\newcommand{\J}[4]{{\sl #1} {\bf #2} (#3) #4}
\newcommand{\andJ}[3]{{\bf #1} (#2) #3}
\newcommand{\AP}{Ann.\ Phys.\ (N.Y.)}
\newcommand{\MPL}{Mod.\ Phys.\ Lett.}
\newcommand{\NP}{Nucl.\ Phys.}
\newcommand{\PL}{Phys.\ Lett.}
\newcommand{\PR}{Phys.\ Rev.}
\newcommand{\PRL}{Phys.\ Rev.\ Lett.}
\newcommand{\PTP}{Prog.\ Theor.\ Phys.}
\newcommand{\hepth}[1]{{\tt hep-th/#1}}
\newcommand{\arxivhep}[1]{{\tt arXiv.org:#1 [hep-th]}}

\end{document}